\documentclass[a4paper]{article}

\usepackage{INTERSPEECH2018}
\usepackage[dvipsnames]{xcolor}
\usepackage{tikz,pgfplots,pgf}
\usepackage{amsmath,graphicx}
\usepackage{mathtools}
\usepackage[utf8]{inputenc} 
\usepackage[T1]{fontenc}    
\usepackage{multirow}
\usepackage{array}
\usepackage{color}
\newcolumntype{P}[1]{>{\centering\arraybackslash}p{#1}}

\usepackage{amssymb}

\def\i{\textbf{i}}
\def\j{\textbf{j}}
\def\k{\textbf{k}}

\def\W{\textbf{W}}

\title{Quaternion Convolutional Neural Networks \\for End-to-End Automatic Speech Recognition}
\name{Titouan Parcollet$^{1,2,4}$, Ying Zhang  $^{2,5}$, Mohamed Morchid$^1$, Chiheb Trabelsi$^{2}$, Georges Linarès$^1$, \\Renato De Mori$^{1,3}$ and Yoshua Bengio${^{2,}}^\dagger$\thanks{${}^\dagger$ CIFAR Senior Fellow}}

\address{
  $^1$Université d'Avignon, LIA, France\\
  $^2$Université de Montréal, MILA, Canada\\
  $^3$McGill University, Montréal, Canada\\
  $^4$Orkis, Aix en provence, France\\
  $^5$Element AI, Montréal, Canada}
\email{titouan.parcollet@alumni.univ-avignon.fr, ying.zhlisa@gmail.com, mohamed.morchid@univ-avignon.fr, chiheb.trabelsi@polymtl.ca, georges.linares@univ-avignon.fr, rdemori@cs.mcgill.ca}

\begin{document}
\maketitle


\begin{abstract}
Recently, the connectionist temporal classification (CTC) model coupled with recurrent (RNN) or convolutional neural networks (CNN), made it easier to train speech recognition systems in an end-to-end fashion. However in real-valued models, time frame components such as mel-filter-bank energies and the cepstral coefficients obtained from them, together with their  first and second order derivatives, are processed as individual elements, while a natural alternative is to process such components as composed entities. We propose to group such elements in the form of quaternions and to process these quaternions using the established quaternion algebra. Quaternion numbers and quaternion neural networks have shown their efficiency to process multidimensional inputs as entities, to encode internal dependencies, and to solve many tasks with less learning parameters than real-valued models. This paper proposes to integrate multiple feature views in quaternion-valued convolutional neural network (QCNN), to be used for sequence-to-sequence mapping with the CTC model. Promising results are reported using simple QCNNs in phoneme recognition experiments with the TIMIT corpus. More precisely, QCNNs obtain a lower phoneme error rate (PER) with less learning parameters than a competing model based on real-valued CNNs.
\end{abstract}
 
\noindent\textbf{Index Terms}: quaternion convolutional neural networks, automatic speech recognition, deep learning


\section{Introduction}  
Recurrent (RNN) and convolutional (CNN) neural networks have improved the performance over hidden Markov models (HMM) combined with gaussian mixtures models (GMMs) in automatic speech recognition (ASR) systems \cite{sak2014long, hinton2012deep,abdel2012applying,mirco2017timit,greff2017lstm} during the last decade.
More recently, end-to-end approaches received a growing interest due to the promising results obtained with connectionist temporal classification (CTC)~\cite{graves2006connectionist} combined with RNNs \cite{sak2014long} or CNNs \cite{zhang2017towards}.\\
However, despite such evolution of models and paradigms, the acoustic features remain almost the same. The main motivation is that filters spaced linearly at low frequencies and logarithmically at high frequencies make it possible to capture phonetically important acoustic correlates. Early evidence was provided in \cite{davis1990comparison} showing that mel frequency scaled cepstral coefficients (MFCCs) are effective in capturing the acoustic information required to recognize syllables in continuous speech.
Motivated by these analysis, a small number of MFCCs (usually $13$) with their first and second time-derivatives, as proposed in \cite{furui1986speaker}, have been found suited for statistical and neural ASR systems. In most systems, a time frame of the speech signal is represented by a vector with real-valued elements that express sequences of MFCCs, or filter energies, and their temporal context features. A concern addressed in this paper, is the fact that the relations between different views of the features associated with a frequency are not explicitly represented in the feature vectors used so far. 
Therefore, this paper proposes to:
\begin{itemize}
\item Introduce a new quaternion representation (Section \ref{sec:qalgebra}) to encode multiple views of a time-frame frequency in which different views are encoded as values of imaginary parts of a hyper-complex number. Thus, vectors of quaternions are embedded using operations defined by a specific quaternion algebra to preserve a distinction between features of each frequency representation.
\item Merge a quaternion convolutional neural network (QCNN, Section \ref{sec:QCNN}) with the CTC in a unified and easily reusable framework\footnote{The full code is available at https://git.io/vx8so}.
\item Compare and evaluate the effectiveness of the proposed QCNN to an equivalent real-valued model on the TIMIT \cite{garofolo1993darpa} phonemes recognition task (Section \ref{sec:exps}). 
\end{itemize}
There are advantages which could derive from bundling groups of numbers into a quaternion. Like capsule networks~\cite{hinton2017capsule}, quaternion networks create a tighter association between small groups of numbers rather than having one homogeneous representation. In addition, this kind of structure reduces the number of required parameters considerably, because only one weight is necessary between two quaternion units, instead of 4$\times 4=16$. The hypothesis tested here is whether these advantages lead to better generalization.
The conducted experiments on the TIMIT dataset yielded a phoneme error rate (PER) of $19.64$\% for QCNNs which is significantly lower than the PER obtained with real-valued CNNs ($20.57$\%), with the same input features. Moreover, from a practical point of view, the resulting networks have a considerably smaller memory footprint due to a smaller set of parameters.


\section{Quaternion algebra}
\label{sec:qalgebra}
The quaternions algebra $\mathbb{H}$ defines operations between quaternion numbers. A quaternion Q is  an extension of a complex number defined in a four dimensional space. $Q = r1 + x\i + y\j + z\textbf{k}$, with, r, x, y, and z four real numbers, and $1$, \i, \j, and \textbf{k} are the quaternion unit basis. Such a definition can be used for describing spatial rotations that can also be represented by the following matrix of real numbers:
\begin{align}
Q = 
\begin{bmatrix}
   r & x & y & z \\
   -x & r & -z & y \\
   -y & z & r & -x \\
   -z & -y & x & r 
\end{bmatrix}.
\end{align}
In a quaternion, $r$ is the real part while $x\i+y\j+z\k$ is the imaginary part ($I$) or the vector part.
Basic quaternion definitions are
\begin{itemize}
\renewcommand{\labelitemi}{$\bullet$}
\item all products of $\i,\j$,$\k$ are: $\i^2=\j^2=\k^2=\i\j\k=-1$,
\item conjugate $Q^*$ of $Q$ is: $Q^*=r1-x\i-y\j-z\k$,
\label{eq:conjugate}
\item unit quaternion $Q^\triangleleft=\frac{Q}{\sqrt{r^2+x^2+y^2+z^2}}$,
\label{eq:normalized}
\label{eq:norm}
\item the Hamilton product $\otimes$ between $Q_1$ and $Q_2$ is defined as follows: 
\begin{align}
Q_1 \otimes Q_2=&(r_1r_2-x_1x_2-y_1y_2-z_1z_2)+\nonumber \\
			&(r_1x_2+x_1r_2+y_1z_2-z_1y_2)\i+\nonumber \\
            &(r_1y_2-x_1z_2+y_1r_2+z_1x_2)\j+\nonumber \\
            &(r_1z_2+x_1y_2-y_1x_2+z_1r_2)\k.
\label{eq:hamilton}
\nonumber
\end{align}
\end{itemize}
The Hamilton product is used in QCNNs to perform transformations of vectors representing quaternions, as well as scaling and interpolation between two rotations following a geodesic over a sphere in the $\mathbb{R}^3$ space as shown in~\cite{minemoto2017feed}.


\section{Quaternion convolutional neural networks}
\label{sec:QCNN}

This section defines the internal quaternion representation (Section \ref{subsec:qinternal}), the quaternion convolution (Section \ref{subsec:qconv}), a proper parameter initialization (Section \ref{subsec:init}), and the connectionist temporal classification (Section \ref{subsec:ctc}).

\subsection{Quaternion internal representation}
\label{subsec:qinternal}
The QCNN is a quaternion extension of well-known real-valued and complex-valued deep convolutional networks (CNN) \cite{he2016deep,chiheb2017complex}. The quaternion algebra is ensured by manipulating matrices of real numbers. Consequently, a traditional $2D$ convolutional layer, with a kernel that contains $N$ feature maps, is split into 4 parts: the first part equal to $r$, the second one to $x\i$, the third one to $y\j$ and the last one to $z\textbf{k}$ of a quaternion $Q = r1+x\i+y\j+z\textbf{k}$. Nonetheless, an important condition to perform backpropagation in either real, complex or quaternion neural networks is to have cost and activation functions that are differentiable with respect to each part of the real, complex or quaternion number. Many activation functions for quaternion have been investigated \cite{xu2017learning} and a quaternion backpropagation algorithm have been proposed in \cite{nitta1995quaternary}. Consequently, the split activation \cite{arena1994neural,parcollet2016quaternion} function is applied to every layer and is defined as follows:
\begin{equation}
\alpha(Q)=\alpha(r)+\alpha(x)\i+\alpha(y)\j+\alpha(z)\textbf{k},
\end{equation}
with $\alpha$ corresponding to any standard activation function.

\subsection{Quaternion-valued convolution}
\label{subsec:qconv}
\begin{figure}[h!]
 \begin{center}
\scalebox{0.46}{
\includegraphics[width=1\textwidth,origin=c]{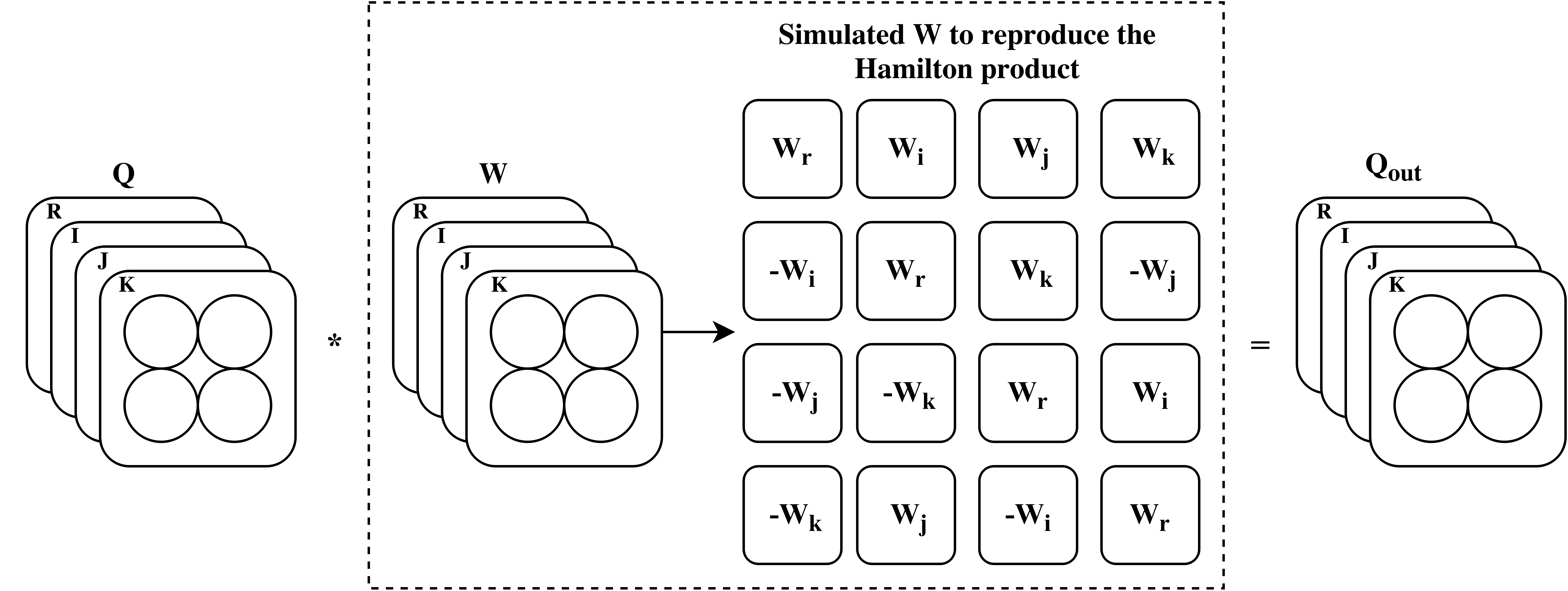}
}
\caption{Illustration of the quaternion convolution}\label{fig:qconv}
\end{center}
\end{figure}

Following a recent proposition for convolution of complex numbers\cite{chiheb2017complex} and quaternions \cite{chase2017quat}, this paper presents basic neural networks convolution operations using quaternion algebra. The convolution process is defined in the real-valued space by convolving a filter matrix with a vector. In a QCNN, the convolution of a quaternion filter matrix with a quaternion vector is performed. For this computation, the Hamilton product is computed using the real-valued matrices representation of quaternions. Let $W =R + X\i + Y\j + Z\textbf{k}$ be a quaternion weight filter matrix, and $X_p = r + x\i + y\j + z\textbf{k}$ the quaternion input vector. The quaternion convolution w.r.t the Hamilton product $W \otimes X_p$ is defined as follows:
\begin{align}
W \otimes X_p = &(Rr-Xx-Yy-Zz)+\nonumber \\
			&(Rx+Xr+Yz-Zy)\i+\nonumber \\
            &(Ry-Xz+Yr+Zx)\j+\nonumber \\
            &(Rz+Xy-Yx+Zr)\textbf{k},
\end{align}
and can thus be expressed in a matrix form:

\begin{equation}
W \otimes X_p =
\begin{bmatrix}
   R & -X & -Y & -Z \\
   X & R & -Z & Y \\
   Y & Z & R & -X \\
   Z & -Y & X & R 
\end{bmatrix}
* 
\begin{bmatrix}
    r \\ x \\ y \\ z 
\end{bmatrix}
=
\begin{bmatrix}
    r' \\ x'\i \\ y'\j \\ z'\textbf{k} 
\end{bmatrix},
\end{equation}
An illustration of such operation is depicted in Figure~\ref{fig:qconv}.

\subsection{Weight initialization}
\label{subsec:init}
Weight initialization is crucial to efficiently train neural networks. An appropriate initialization improves training speed and reduces the risk of exploding or vanishing gradient. A quaternion initialization is composed of two steps. First, for each weight to be initialized, a purely imaginary quaternion $q_{imag}$ is generated following an uniform distribution in the interval $[0,1]$. The imaginary unit is then normalized to obtain $q_{imag}^\triangleleft$ following the quaternion normalization equation. The later is used alongside to other well known initializing criterion such as \cite{glorot2010understanding} or \cite{he2015delving} to complete the initialization process of a given quaternion weight named $w$. Moreover, the generated weight has a polar form defined by : 
\begin{align}
\centering
w=|w|e^{\textbf{n}\theta}=|w|(cos(\theta) + \textbf{n}sin(\theta)),
\end{align}
 with
 \begin{align}
 \textbf{n} = \dfrac{x\i + y\j + z\textbf{k}}{|w|sin(\theta)}.
 \end{align}
 Therefore, $w$ is generated as follows:
\begin{itemize}
	\item $ w_\textbf{r} = \phi * q^\triangleleft_{imag\textbf{r}} * cos(\theta)$,
	\item $ w_\textbf{i} = \phi * q^\triangleleft_{imag\textbf{i}} * sin(\theta)$,
	\item $ w_\textbf{j} = \phi * q^\triangleleft_{imag\textbf{j}} * sin(\theta)$,
	\item $ w_\textbf{k} = \phi * q^\triangleleft_{imag\textbf{k}} * sin(\theta)$.
\end{itemize}
However, $\phi$ represents a randomly generated variable with respect to the variance of the quaternion weight and the selected initialization criterion. The initialization process follows \cite{glorot2010understanding} and \cite{he2015delving} to derive the variance of the quaternion-valued weight parameters. Therefore, the variance of $\W$ has to be investigated:
\begin{align}
	Var(\W) = \mathbb{E}(|\W|^2) -  [\mathbb{E}(|\W|)]^2.
\end{align}
$[\mathbb{E}(|\W|)]^2$ is equals to $0$ since the weight distribution is symmetric around $0$. Nonetheless, the value of $Var(\W) = \mathbb{E}(|\W|^2)$ is not trivial in the case of quaternion-valued matrices. Indeed, $W$ follows a Chi-distributed with four degrees of freedom (DOFs) and $Var(\W) = \mathbb{E}(|\W|^2)$ is expressed and computed as follows:
\begin{align}
	Var(\W) = \mathbb{E}(|\W|^2) =\int_{0}^{\infty} x^2f(x) \, \mathrm{d}x=4\sigma^2.
\end{align}
Therefore, in order to respect the He Criterion \cite{he2015delving}, the variance would be equal to:
\begin{align}
\sigma = \frac{1}{\sqrt[]{2(n_{in})}}.
\end{align}

\subsection{Connectionist Temporal Classification}
\label{subsec:ctc}
In the acoustic modeling part of ASR systems, the task of sequence-to-sequence mapping from an input acoustic signal $X=[x_1,...,x_n]$ to a sequence of symbols $T=[t_1,...,t_m]$ is complex due to:
\begin{itemize}
\item $X$ and $T$ could be in arbitrary length.
\item The alignment between $X$ and $T$ is unknown in most cases.
\end{itemize}
Specially, $T$ is usually shorter than $X$ in terms of phoneme symbols.

To alleviate these problems, connectionist temporal classification (CTC) has been proposed \cite{graves2006connectionist}. First, a softmax is applied at each timestep, or frame, providing a probability of emitting each symbol $X$ at that timestep. This probability results in a symbol sequences representation $P(O|X)$, with $O = [o_1,...,o_n]$ in the latent space $O$. A blank symbol $'-'$ is introduced as an extra label to allow the classifier to deal with the unknown alignment. Then, $O$ is transformed to the final output sequence with a many-to-one function $g(O)$ defined as follows:
\begin{align}
\left.
    \begin{array}{ll}
        g(z_1,z_2,-,z_3,-) \\
        g(z_1,z_2,z_3,z_3,-) \\
        g(z_1,-,z_2,z_3,z_3) \\
    \end{array}
\right \}=(z_1,z_2,z_3). 
\end{align}
Consequently, the output sequence is a summation over the probability of all possible alignments between $X$ and $T$ after applying the function $g(O)$. Accordingly to \cite{graves2006connectionist} the parameters of the models are learned based on the cross entropy loss function: 
\begin{align}
 \sum\nolimits_{X, T \in train}-\log(P(O|X)). 
\end{align}
During the inference, a best path decoding algorithm is performed. Therefore, the latent sequence with the highest probability is obtained by performing argmax of the softmax output at each timestep. The final sequence is obtained by applying the function $g(.)$ to the latent sequence.


\section{Experiments}
\label{sec:exps}
The performance and efficiency of the proposed QCNNs is evaluated on a phoneme recognition task. This section provides details on the dataset and the quaternion features representation (Section \ref{sec:dataset}), the models configurations (Section \ref{sec:model}), and finally a discussion of the observed results (Section \ref{sec:results}).

\subsection{TIMIT dataset and acoustic features of quaternions}
\label{sec:dataset}
The TIMIT \cite{garofolo1993darpa} dataset is composed of a standard 462-speaker training dataset, a 50-speakers development dataset and a core test dataset of $192$ sentences. During the experiments, the SA records of the training set are removed and the development set is used for early stopping. The raw audio is transformed into $40$-dimensional log mel-filter-bank coefficients with deltas, delta-deltas, and energy terms, resulting in a one dimensional vector of length $123$. An acoustic quaternion $Q(f,t)$ associated with a frequency $f$ and a time frame $t$ is defined as follows:
\begin{align}
Q(f,t) = 0 + e(f,t)\i + \frac{\partial e(f,t)}{\partial t} \j + \frac{\partial^2 e(f,t)}{\partial^2 t} \k.
\end{align}
It represents multiple views of a frequency $f$ at time frame $t$, consisting of the energy $e(f,t)$ in the filter band corresponding to $f$, its first time derivative describing a slope view, and its second time derivative describing a concavity view. Finally, a unique quaternion is composed with the three corresponding energy terms. Thus, the quaternion input vector length is $41$ ($\frac{123}{3}$).

\subsection{Models architectures}
\label{sec:model}
The architectures of both CNN and QCNN models are inspired by~\cite{zhang2017towards}. A first $2$D convolutional layer is followed by a maxpooling layer along the frequency axis. Then, $n$ $2$D convolutional layers are included, together with $3$ dense layers of sizes $1024$ and $256$ respectively for real- and quaternion-valued models (with $n\in[6,10]$). Indeed, the output of a dense quaternion-valued layer has $256 \times 4 = 1024$ nodes and is $4$ times larger than the number of units.
The filter size is rectangular $(3,5)$, and a padding is applied to keep the sequence and signal sizes unaltered. The number of feature maps varies from $32$ to $256$ for the real-valued models and from $8$ to $64$ for quaternion-valued models. Indeed, the number of output feature maps is $4$ times larger in the QCNN due to the quaternion convolution, meaning $32$ quaternion-valued feature maps correspond to $128$ real-valued ones. The PReLU activation function is employed for both models~\cite{he2015delving}. A dropout of $0.3$ and a $L_2$ regularization of $1e^{-5}$ are used across all the layers, except the input and output ones. CNNs and QCNNs are trained with the Adam learning rate optimizer and vanilla hyperparameters~\cite{kingma2014adam} during $100$ epochs. Then, a fine-tuning process of $50$ epochs is performed with a standard $sgd$ and a learning rate of $1e^{-5}$. Finally, the standard CTC loss function defined in~\cite{graves2006connectionist} and implemented in \cite{chollet2015keras} is applied. Experiments are performed on Tesla P100 and Geforce Titan X GPUs.

\subsection{Results and discussion}
\label{sec:results}
Results on the phoneme recognition task of the TIMIT dataset are reported in Table \ref{table:results}. It is worth noticing the important difference in terms of the number of learning parameters between real and quaternion valued CNNs. It is easily explained by the quaternion algebra. In the case of a dense layer with $1,024$ input values and $1,024$ hidden units, a real-valued model will have $1,024^2\approx1$M parameters, while to maintain equal input and output nodes ($1,024$) the quaternion equivalent has $256$ quaternions inputs and $256$ quaternion-valued hidden units. Therefore the number of parameters for the quaternion model is $256^2\times4\approx0.26$M. Such a complexity reduction turns out to produce better results and may have other advantages such as a smallest memory footprint while saving NN models. Moreover, the reduction of the number of parameters does not result in poor performance in the QCNN. Indeed, the best PER reported is $19.64$\% from a QCNN with $256$ feature maps and $10$ layers, compared to a PER of $20.57$\% for a real-valued CNN with $64$ feature maps and $10$ layers. It is worth underlying that both model accuracies are increasing with the size and the depth of the neural network. However, bigger real-valued feature maps leads to overfitting. In fact, as shown in Table \ref{table:results}, the best PER for a real-valued model is reached with $64$ ($20.57$) feature maps and decreasing at $128$ ($20.62$\%) and $256$ ($21.23$). The QCNN does not suffer from such weaknesses due to the smaller density of the neural network and achieved a constant PER improvement alongside with the increasing number of feature maps. Furthermore, QCNNs always performed better than CNNs independently of the model topologies.   

\begin{table}[ht]
\caption{Experiment results expressed in term of phoneme error rate (PER) percentage of both QCNN and CNN based models on the TIMIT phoneme recognition task. The results are from a $3$ folds average. 'L' stands for number of Layers, 'FM' for number of feature maps, and 'Params' for number of learning parameters. The latter is expressed in order to be equivalent for both models. Therefore, $32$FM is equal to $32$FM for real numbers and $8$ quaternion-valued FM}
\label{table:results}
\begin{center}
\scalebox{0.92}{
\begin{tabular}{ P{3cm}P{1.3cm}P{1.3cm}P{1cm}}
    \hline\hline
    \textbf{Models} & \textbf{Dev PER \%} & \textbf{Test PER \%} & \textbf{Params}\\
    \hline
    $\mathbb{R}$-CNN-6L-32FM & 22.18 & 23.54 & 3.3M\\
    $\mathbb{H}$-QCNN-6L-32FM & 22.16 & 23.20 & 0.87M\\
    $\mathbb{R}$-CNN-10L-32FM & 21.77 & 23.43 & 3.4M\\
    $\mathbb{H}$-QCNN-10L-32FM &22.25 & 23.23 & 0.9M\\
   	\hline
    $\mathbb{R}$-CNN-6L-64FM & 21.19 & 22.12 & 4.8M\\
    $\mathbb{H}$-QCNN-6L-64FM & 21.44 & 21.99 & 1.2M\\
    $\mathbb{R}$\textbf{-CNN-10L-64FM} & 19.53 & \textbf{20,57} & 5.4M\\
    $\mathbb{H}$-QCNN-10L-64FM & 19.78 & 20.44 & 1.4M\\
    \hline
    $\mathbb{R}$-CNN-6L-128FM & 20.33 & 22.14 & 9M\\
    $\mathbb{H}$-QCNN-6L-128FM & 20.12 & 21.33 & 2.3M\\
    $\mathbb{R}$-CNN-10L-128FM & 19.37 & 20.62 & 11.5M\\
    $\mathbb{H}$-QCNN-10L-128FM & 19.02 & 19.87 & 2.9M\\
    \hline
    $\mathbb{R}$-CNN-6L-256FM & 20.43 & 22.25 & 22.3M\\
    $\mathbb{H}$-QCNN-6L-256FM & 19.94 & 20.54 & 5.6M\\
    $\mathbb{R}$-CNN-10L-256FM & 18.89 & 21.23 & 32.1M\\   
    $\mathbb{H}$\textbf{-QCNN-10L-256FM} & 18.33 & \textbf{19.64} & 8.1M\\
    \hline
\end{tabular}
}
\end{center}
\label{tab:multicol}
\end{table}

With much fewer learning parameters for a given architecture, the QCNN performs always better than the real-valued one on the reported task. In terms of PER, an average relative gain of $3.25$\% (w.r.t CNNs result) is obtained on the testing set. It is also worth recalling that the best PER of $19.64$\% is obtained with just a QCNN without HMMs, RNNs, attention mechanisms, batch normalization, phoneme language model, acoustic data normalization or adaptation. Further improvements can be obtained with exactly the same QCNN by just introducing a new acoustic feature in the real part of the quaternions.


\section{Related work}
Early attempts to perform phoneme and phonetic feature recognition with multilayer perceptrons (MLP) were proposed in~\cite{bourlard1996mew,robinson1994application,bengio1992global}. A PER of $26.1$\% is reported in~\cite{robinson1994application} using RNNs. More recently, in \cite{dahl2010phone} a Mean-Covariance Restricted Boltzmann Machine (RBM) is used for recognizing phonemes in the TIMIT corpus using RBM for feature extraction. Along this line of research, in \cite{graves2006connectionist} an approach called the Connectionist Temporal Classification (CTC) has been developed and can be used without an explicit input-output alignment. 
Bidirectional RNNs (BRNNs) are used in \cite{graves2013speech} for processing input data in both directions with two separate hidden layers, which are then composed in an output layer. With standard mel frequency energies, first and second time derivatives a PER of $17.7$\% was obtained. Other recent results with real-valued vectors of similar features are reported in \cite{chorowski2014end,mirco2017timit,lu2016segmental,lu2017multi}. Other types of quaternion valued neural networks (QNNs) were introduced for encoding RGB color relations in image pixels \cite{hsiao2014edge,chen2015color,garg2017vector}, and for classifying human/human conversation topics \cite{parcollet2017deep,parcollet2017quaternion,parcollet2016quaternion}. A quaternion deep convolutional and residual neural network proposed in \cite{chase2017quat} have shown impressive results on the CIFAR images classification task. However, a specific quaternion is used for each RGB color value as in \cite{chiheb2017complex} rather than integrating pixel multiple views as in \cite{kusamichi2004new}, and suggested in this paper for an ASR task.


\section{Conclusions}
\textbf{Summary}.
This paper proposes to integrate multiple acoustic feature views with quaternion hyper complex numbers, and to process these features with a convolutional neural network of quaternions. The phoneme recognition experiments have shown that: 1) Given an equivalent architecture, QCNNs always outperform CNNs with significantly less parameters; 2) QCNNs obtain better results than CNNs with a similar number of learning parameters; 3) The best result obtained with QCNNs is better than the one observed with the real-valued counterpart. This demonstrates the initial intuition that the capability of the Hamilton product to learn internal latent relations helps quaternions-valued neural networks to achieve better results .\\   
\textbf{Limitations and Future Work}.
So far, traditional acoustic features, such as mel filter bank energies, first and second derivatives have shown that significantly good results can be obtained with a relative small set of input features for a speech time frame. Nevertheless, speech science has shown that other multi-view context-dependent acoustic relations characterize signals of phonemes in context.
Future work will attempt to characterize those multi-view features that mostly contribute to reduce ambiguities in representing phoneme events. Furthermore, quaternions-valued RNNs will also be investigated to see if they can contribute to the improvement of recently achieved top of the line results with real number RNNs.


\section{Acknowledgements}
The experiments were conducted using Keras \cite{chollet2015keras}. The authors would like to acknowledge the computing support of Compute Canada and the founding support of Orkis, NSERC, Samsung, IBM and CHIST-ERA/FRQ. The authors would like to thank Kyle Kastner and Mirco Ravanelli for their helpful comments.   

\newpage

\bibliographystyle{IEEEtran}

\bibliography{strings}

\end{document}